# Analyze Unstructured Data Patterns for Conceptual Representation

Aboubakr Aqle, Dena Al-Thani
Information & Computing Technology
College of Science & Engineering, Hamad Bin Khalifa University
Doha, Qatar
{aaqle, dalthani}@hbku.edu.qa

Ali Jaoua
Computer Science & Engineering
College of Engineering, Qatar University
Doha, Qatar
jaoua@qu.edu.qa

*Abstract*—Online news media provides aggregated news and stories from different sources all over the world and up-to-date news coverage. The main goal of this study is to find a solution that is considered as a homogeneous source for the news and to represent the news in a new conceptual framework. Furthermore, the user can easily and quickly find different updated news in a fast way through the designed interface. The Mobile App implementation is based on modeling the multi-level conceptual analysis frame. Discovering main concepts of any domain is captured from the hidden unstructured data that are analyzed by the proposed solution. Concepts are discovered through analyzing data patterns to be structured into a tree-based interface for easy navigation for the end user. Our final experiment results show that analyzing the news before displaying to the end-user and restructuring the final output in a conceptual multilevel structure produces a new display frame for the end user to find the related information of interest.

## I. INTRODUCTION

News-Application Programming Interface (API) is a simple and easy-to-use tool that can return its response for the metadata of the live news headlines. This news comes from different range of sources and blogs that are published on press and media. Sources can vary in their specialized domains such as politics, sports, or economics.

These APIs can be used to display live news headlines for the end user in a fast and easy structure. However, News can be retrieved and displayed on smart-phones and small screens in general, by a new different approach. The major idea behind the proposed approach is to create a simple look and feel design that will enhance the user experience (UX).

This paper aim is to show the impact of using formal concept analysis for conceptual modeling, concepts discovery, and extraction from different unstructured data patterns. We will illustrate how we can take advantage of online news APIs to display extracted news in a different manner that can serve different types of end users. The new approach will help users to find their target news from different online sources with a new analysis and presentation design that can address the users' needs with minimal interaction, which improves the user experience.

Most press and media data are accessible through direct online websites or through authorized APIs. Extracting information from such hidden databases can be achieved using these APIs as they connect to different sources and retrieve the news as an input of unstructured data. Consequently, the proposed system will have tree-based graphical representation as a final result on the browser interface for the end user.

Accessing different online news websites to have an overview for the latest news that is important to the end user is effort and time consuming. The user will need to reach these websites and to read the different news headlines to select the ones which most closely match user interest before proceeding to read these articles. In order to overcome this gap, our research proposes a Mobile App that has an integrated, unified, and interactive interface for multiple different online news and media websites. This paper focuses on using conceptual modeling and discovery algorithm with concepts extraction to appoint the available news along with their features automatically. This methodology of data analysis and modeling provides a multilevel browsing environment that enhances the user options for reaching the target news from different sources with the minimum effort.

The paper is divided into five further sections: section II deals with background and related work overview, section III includes the proposed solution for the conceptual browser tree for the news and media websites, experimental results are discussed in section IV; the conclusion is reported in section V; and finally, the future work is presented in section VI.

## II. BACKGROUND AND RELATED WORK

### A. Basic Concepts of Binary Relation and Maximal Gain

Let objects set denoted by **O**, and properties set denoted by **P**. Binary relation denoted as **R** is a Cartesian product subset of objects and properties (**O×P**) where the element of R is (x, y).

Non-empty rectangular relation or rectangle denoted by **RE** of the binary relation **R** is any Cartesian product **A×B** included in **R** where, **A** ⊆ **O** and **B** ⊆ **P** and both of **A** and **B** ≠ ∅.

**A** is the domain Dom(**R**), while **B** is the codomain Cod(**R**) of the rectangle. ‖**A**‖ is the cardinality of **A**. The gain of a given rectangle **RE** that is (**A** × **B**) is calculated by equation (1):

$$g(\mathbf{RE}) = ( \|\mathbf{A}\| \times \|\mathbf{B}\| ) - ( \|\mathbf{A}\| + \|\mathbf{B}\| ) \quad (1)$$

A rectangular relation **RE** of **A** × **B** containing an element that is (**a**, **b**) of a binary relation **R** is considered as optimal if and only if it realizes that maximal gain g(**RE**) among all rectangular relations containing the element (**a**, **b**). The Optimal rectangle is representing the important data associations that involving the element of (**a**, **b**) [1].

Relation R has coverage denoted as CV for a set of rectangles that is CV = (RE$_1$, RE$_2$... RE$_n$), where any element (**a**, **b**) of the relation R to be at least in one rectangle of the whole coverage CV.

### B. Formal Concept Analysis

Formal concept analysis (FCA) is a defined mathematical framework for data analysis by determining the relations between a specific objects' set and attributes' set [2]. FCA has many terminologies such as, formal context and formal concept.

Formal context is triple **K**= (**O**, **P**, **I**) where **O** is a finite set of objects that represent tuples in database instance (DBI), **P** is a finite set properties or attributes that represent columns in DBI, **I** is the relation between **O** and **P** [3]. Table I is a formal context example that has a set of objects **O** = {O$_1$, O$_2$, O$_3$}, set of properties **P** = {P$_1$, P$_2$, P$_3$}, and has element (O$_1$, P$_1$) ∈ I, but element (O$_2$, P$_3$) ∉ I.

Formal concept of the context K is a pair (**X**, **Y**), such that **X** ⊆ **O**, **Y** ⊆ **A**, such X' = Y and Y' = X [3], [4]. For example, from Table I, objects {o1, o2} have common properties {P$_1$, P$_2$} and for this attributes set, they share the same initial objects set of {O$_1$, O$_2$}, so ({O$_1$, O$_2$}, {P$_1$, P$_2$}) is a formal concept FC-1. We can have full coverage of the relation by extracting the second formal concept of ({O$_3$}, {P$_2$, P$_3$}) as FC-2 to have the all formal concepts that cover the relation to be: FC-1 and FC-2.

TABLE I. FORMAL CONTEXT

| Objects | P$_1$ | P$_2$ | P$_3$ | |
|---|---|---|---|---|
| O$_1$ | 1 | 1 | | FC-1 |
| O$_2$ | 1 | 1 | | |
| O$_3$ | | 1 | 1 | FC-2 |

### C. Related Work

Automated text summarization is considered as the major task for processing huge amount of data available on the web. Many studies have been done for this area [6-9].

One of the techniques that can be used for text summarization is formal concept analysis, which defines methods for data clustering through characterized patterns. Formal concept associated pattern specified a set of objects that are sharing the maximum set of properties or attributes in the context.

Alja'am et al. [5] proposed a method for selecting optimal concepts to minimize information representation, which is a text summarization process based on word similarity method. The idea is to extract a summary from document and association between contained keywords in the document. First, we are creating a binary context where objects are sentences and attributes are words. List of empty words is provided to remove these words from the text. Then from this binary context, we find optimal concept for all sentences to generate the summary keywords. Decomposition process depends on the maximum weight of the words associated to the text.

The proposed method considered as unsupervised classifier to generate summary keywords for the domain text, which is news for this paper. The outcome keywords can be arranged in a multilevel tree-structured design to be known as news conceptual browser.

### III. PROPOSED SOLUTION

The proposed solution provides Mobile App for structuring daily general news from different online news media agencies (News Media-1 as NM1, News Media-2 as NM2 and News Media-3 as NM3). Our conceptual browser methodology shown at figure-1 as the below steps:

### A. Retrieve News by APIs

Hidden data are important factor for data analysis process that need to be accessed by authorized APIs. We registered different three APIs for this paper as NM1, NM2 and NM3. That allow us to send a request for the updated top news, and get a response in JSON (JavaScript Object Notation) format. JSON used to transmit data objects consisting of array data types that containing attribute–value pairs of the published news stores.

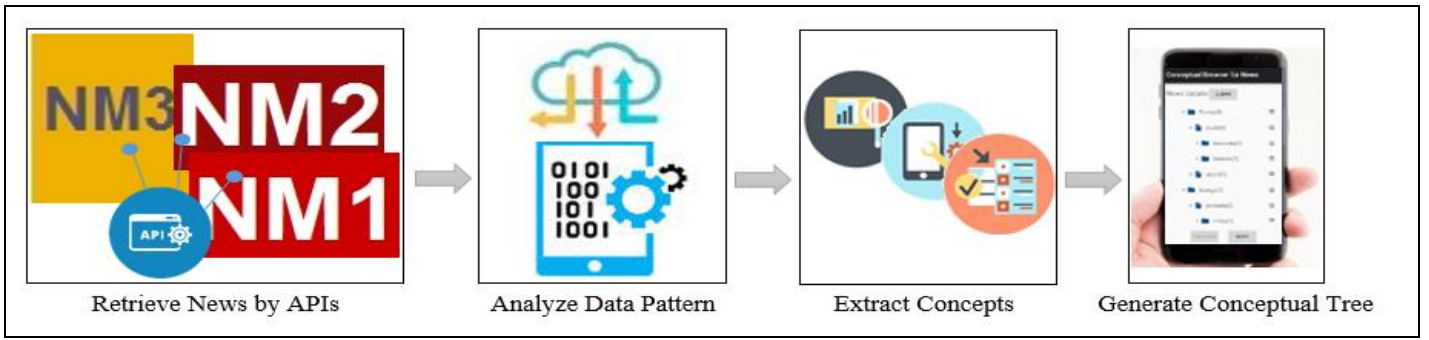

Fig. 1. Proposed Solution Methodology

## B. Analyze Data Pattern

Data extraction process is handling responses from different news media, which need to be unified to one single structure for the analysis process of the data pattern at the Mobile App. The unified structure is considered as database instance from online news media databases.

Static data analysis leads us to three fields/attributes that define each news object. The three mapped unified attributes to our solution are: title for the news article headline, description for the article details, and finally the URL address as a link to the article.

Preprocessing the data at initial stage to analyze the data pattern for the title field as the following two major steps:

1. Natural Language Processing (NLP): applied to article headline to analyze data and prevent duplicated estimation for similar keywords to be extracted as follows:

   i. Stop words removal: filtering input text and removing unnecessary stop words reduces keywords for processing.

   ii. Stemming: converting different words into their root words by removing suffixes or affixes that will reduce the number of terms and the complexity of the data. Porter stemming algorithm is applied for English documents.

   iii. Tokenization: splitting text into sentences/words to be used for indexing stage. Text is divided into sentences based on punctuation marks/blank spaces, and then sentences will be divided into words to form a matrix.

2. Indexing: fundamental step for construction binary relation, where weighting values are calculated for words. The relation is between sentences as objects and words as properties. The weight is calculated by using relative Term Frequency (**tf**) that determines the importance of every term within the document given by equation (2):

$$tf_{ij} = \frac{n_{ij}}{\sum_k n_{kj}} \quad (2)$$

- $n_{ij}$: number of occurrences of term i in document j.
- $\sum_k n_{kj}$: number of all terms in document j.

## C. Extract Concepts and Generate Concpetual Tree

The conceptual browser process after the data analysis stage is processing with the final stage of generating optimal concepts as follows:

1) Create a binary formal context where objects are sentences and attributes are keywords.
2) Find optimal concepts from the binary context for all sentences to generate summary keywords.
3) Assign keyword that has the maximum weight for each optimal concept.
4) Finally, organize keywords that represent the concepts into a heap with greater weight appear at a higher top level of the tree.

Each concept is labeled with a significant keyword that has maximum weight (equation 2) selected from concept coverage.

## IV. EXPERIMENTAL RESULTS

Our experiment will be depending on walkthrough testing for finding certain news that matter the end user on the Mobile Apps of the news media. It will then compare how we can find the same news on our proposed conceptual browser Mobile App. In the first case, we assume the user is looking for latest press statements of the previous US president George W. Bush. The second showcase, the user is interested in Catalonia crisis.

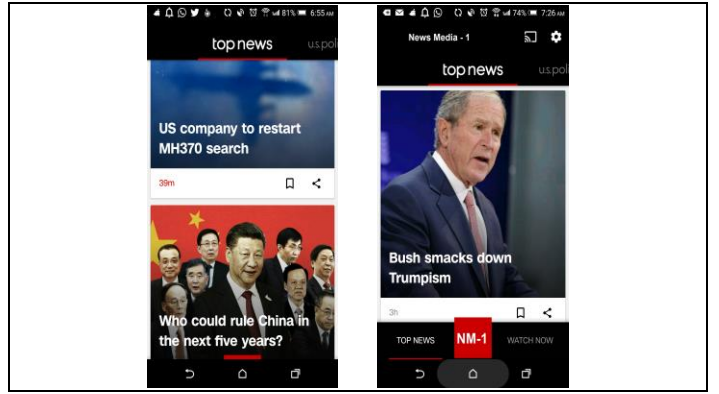

Fig. 2. News Media Agency-1 (NM1) Mobile App

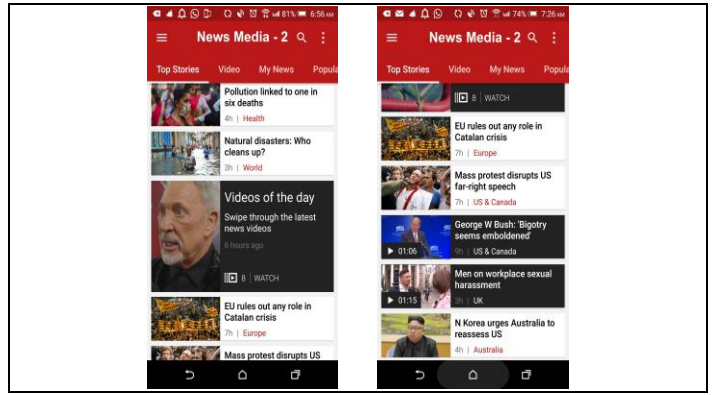

Fig. 3. News Media Agency-2 (NM2) Mobile App

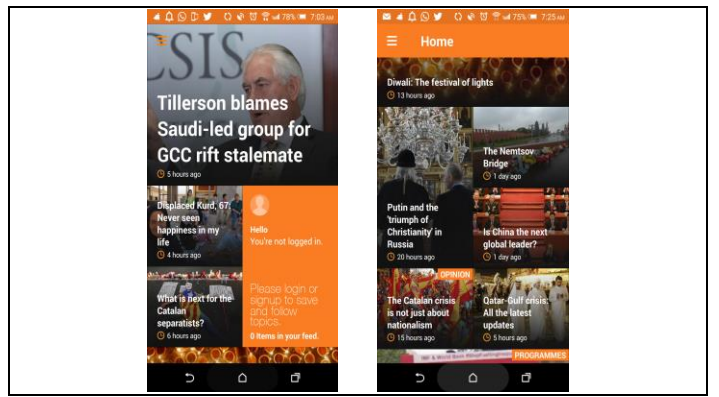

Fig. 4. News Media Agency-3 (NM3) Mobile App

We can notice from figures 2, 3 and 4 that user will scroll down and reads all articles' headings till finding targeted news to navigate to it and reads the details. That was the case for news media agencies NM1 and NM2 for Bush articles and news media agency NM3 for Catalan crisis article.

While our proposed Mobile App for news conceptual browser, display all discovered concepts at first screen, which enables the user to find the targeted news without scrolling down or reading many articles' headlines. Thus, the proposed Mobile App will save the users time and effort during the search process. The first screen of the integrated Mobile App displays the top news organized in a tree structure view for the extracted concepts as well as the number of the matched news between brackets. The second screen designed as view-list, which displays the chosen concept/keyword related articles' description.

Walkthrough testing scenario showed that the proposed Mobile App as news conceptual browser achieved in satisfying the end user demand. The experiment showed that process also save users time in searching for the news they are interested in as shown in figure 5 for Bush articles and figure 7 for Catalan crisis articles:

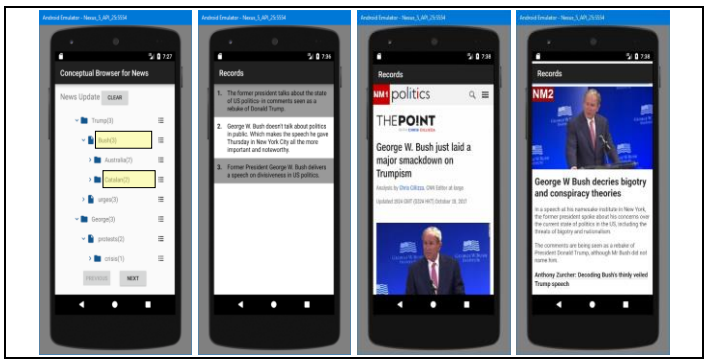

Fig. 5. All Bush Articles with Single Click from Different Sources

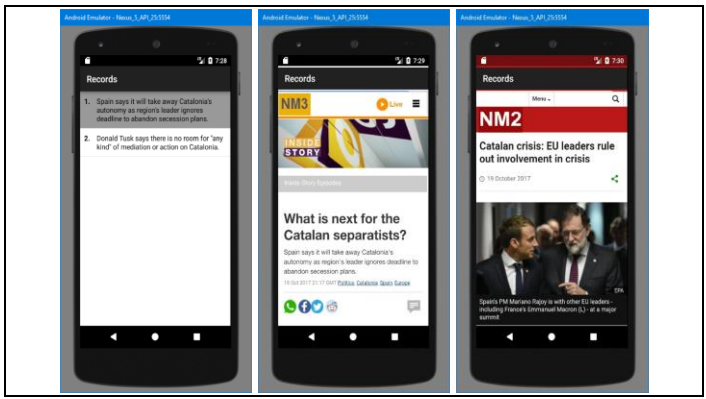

Fig. 6. All Catalan Crisis Articles with Single Click from Different Sources

## V. CONCLUSION

The system utilizes formal concept analysis for conceptual clustering and knowledge discovery. Finding optimal concepts is a heuristic process for determining the domain coverage.

The proposed Mobile App solution is domain independent, multi-source, and presents a discovering framework, which can be integrated with news media networks. The solution is based on concepts extraction and conceptual browser methodologies, for the retrieved general news. The purpose is to make a new representational structure for the extracted information to the end user that facilitates the news reading.

The screen design of the retrieved news as a multi-level tree helps to shorten the query path, and automatically reformulates the query based on the user clicks on the extracted concepts. Conceptual browser methodology will decrease the user interaction with the system, thus allowing user to reach the targeted information in the shortest time possible.

## VI. FUTURE WORK

This paper has demonstrated the potential of the conceptual browser model for decreasing user interaction with the system, and subsequent user satisfaction. New opportunity for extending the scope of this paper appears. One direction for future research is to support the visually impaired (VI) users access the web with minimal interaction when searching for the latest news that is relevant to their domain of interest. This particular avenue for research is potential as many previous studies have highlighted the challenges that VI users encounter when reviewing large amount of information [10]. Another direction for VI users is for the search engine's return results to be presented in a conceptual browser tree. The tree nodes can be accompanied by snippets that are considered as a summary for the concepts that lead the VI user to content.

ACKNOWLEDGMENT

This contribution was made possible by NPRP grant No. 07-794-1-145 and GSRA grant No. 04-1-0514-17066 from the Qatar National Research Fund (a member of Qatar Foundation). The statements made herein are solely the responsibility of the authors.